\documentclass[runningheads]{llncs}
\usepackage{xurl}
\usepackage{float}

\usepackage[T1]{fontenc}
\usepackage{graphicx}
\begin{document}
\title{A user-friendly SPARQL query editor powered by lightweight metadata}
%
%\titlerunning{Abbreviated paper title}
% If the paper title is too long for the running head, you can set
% an abbreviated paper title here
%
\author{Vincent Emonet\orcidID{0000-0002-1501-1082} \and
Ana-Claudia Sima\orcidID{0000-0003-3213-4495} \and
Tarcisio {Mendes de Farias}\orcidID{0000-0002-3175-5372}}
\authorrunning{V. Emonet et al.}
% First names are abbreviated in the running head.
% If there are more than two authors, 'et al.' is used.
%
\institute{SIB Swiss Institute of Bioinformatics, Lausanne, Switzerland \\
\email{\{vincent.emonet, ana-claudia.sima, tarcisio.mendes\}@sib.swiss}}
\maketitle              % typeset the header of the contribution
\begin{abstract}
SPARQL query editors often lack intuitive interfaces to aid SPARQL-savvy users to write queries. %These editors are either proprietary, limited to specific triple stores, or lack data-aware autocomplete features. 
To address this issue, we propose an easy-to-deploy, triple store-agnostic and open-source query editor that offers three main features: (i) automatic query example rendering, (ii) precise autocomplete based on existing triple patterns including within \texttt{SERVICE} clauses, and (iii) a data-aware schema visualization. It can be easily set up with a custom HTML element. The tool has been successfully tested on various public endpoints, and is deployed online at \url{https://sib-swiss.github.io/sparql-editor} with open-source code available at \url{https://github.com/sib-swiss/sparql-editor}.

\end{abstract}
\section{Introduction}
The Semantic Web of data has been growing considerably in recent decades through the deployment of public query endpoints on the Web. For example, \url{yummydata.org} catalogues more than 55 SPARQL endpoints that are of interest to the biomedical community. However, SPARQL endpoints often lack an intuitive Web-based interface that effectively helps SPARQL-savvy users to write queries, notably missing autocomplete \cite{jiang2024autocomplete}. 
Several triple stores such as Stardog\footnote{\url{https://www.stardog.com}} and GraphDB\footnote{\url{https://graphdb.ontotext.com}} propose query editors but they are proprietary and triple store-dependent solutions, or they do not have an autocomplete solution based on a data-aware schema (i.e., a data schema built on the existing data). Alternatively, the \textit{Qlever UI query editor} is open source\footnote{\url{https://github.com/ad-freiburg/qlever-ui}} and provides autocomplete, but works only on the Qlever triple store. In \cite{bast2022efficient}, authors describe the Qlever autocomplete methodology that requires to send an SPARQL query for each autocomplete request. As a result, if this methodology is implemented over non-Qlever endpoints, each autocomplete request can take several seconds compromising usability and increasing the endpoint server load.  
% \cite{bast2017qlever}

Other open-source query editors exist; however, they are tailored to a specific dataset (e.g., \textit{Wikidata Query Service} \cite{10.1145/3543873.3585579}) or lack basic relevant features, such as autocomplete (e.g., \textit{LODEstar}\footnote{\url{https://github.com/EBISPOT/lodestar}} - additionally, the latter has not been maintained since 2018). Currently, a widely used open-source SPARQL editor is  \textit{YASGUI} \cite{rietveld2016yasgui} that is used by platforms such as GraphDB, Oxigraph and UniProt\footnote{\url{https://sparql.uniprot.org}}. Although YASGUI offers user-friendly features like syntax highlighting, customizable autocomplete, endpoint selection, and modular results rendering, it lacks automatic adaptation to specific SPARQL endpoints. Its autocomplete suggestions are generic and unaware of the query context, requiring manual configuration for each endpoint. To the best of our knowledge, no existing query editor supports autocomplete within \texttt{SERVICE} clauses (i.e., federated queries).

To address these issues, we propose an SPARQL query editor based on YASGUI that is easy to deploy, open-source and triple store agnostic, dedicated to helping users write queries over Resource Description Framework (RDF) data. The editor automatically loads and renders query examples from any SPARQL endpoint, and provides a context-aware autocomplete feature.

%There is a work in progress by the Qlever team to write a Language Server Protocol for SPARQL that would enable advanced features for SPARQL editing in most IDEs (e.g. VSCode, Neovim) in addition to web browsers: https://github.com/IoannisNezis/Qlue-ls

\section{Implementation}\label{sec2}

To support users to write queries for a given SPARQL endpoint, we extended the \textit{Yasgui} environment \cite{rietveld2016yasgui} with three main features: (i) a query examples section, automatically constructed from endpoint metadata; (ii) precise autocomplete that adapts according to the SPARQL statements written by the user; and (iii) a data-aware schema visualization. To facilitate deployment, this extension is provided as an npm package\footnote{\url{https://www.npmjs.com/package/@sib-swiss/sparql-editor}} that can be easily set up by any SPARQL endpoint maintainer using a single custom HTML element \texttt{<sparql-editor/>}.  

\subsection{Displaying query examples}

\begin{figure}[h!]
\includegraphics[width=\textwidth]{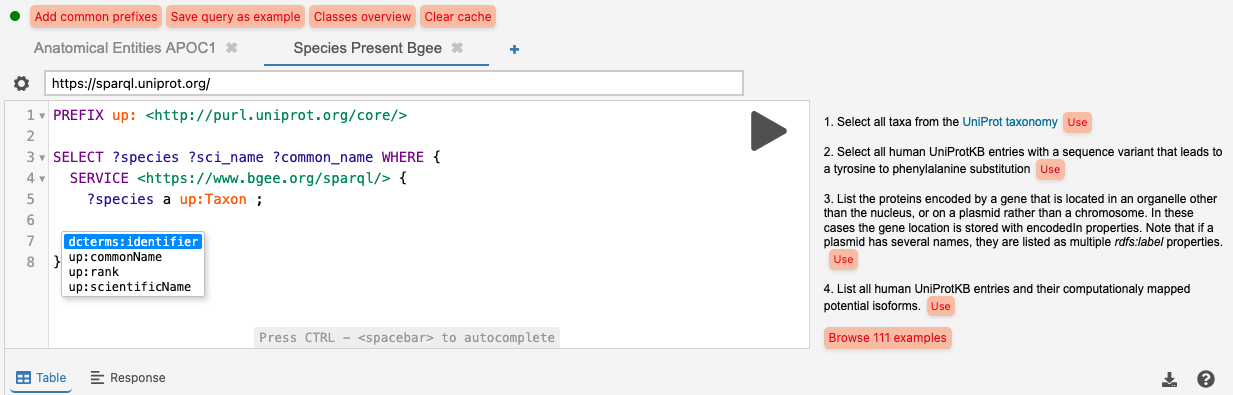}
\caption{Our query editor showing on the right side example queries stored as metadata and fetched from a SPARQL endpoint. By pressing CTRL+Space after \texttt{?species}, the autocomplete suggests only predicates which are asserted to instances of \texttt{up:Taxon}.} \label{fig1}
\end{figure}

Our query editor automatically pulls SPARQL query examples stored as metadata in any given triple store and displays them to the user. To enable this, the SPARQL query examples should be described using the lightweight metadata schema proposed in \cite{gigascience-paper}. In this methodology, SPARQL queries and their descriptions are defined mostly by reusing existing vocabularies such as the W3C recommendation Shapes Constraint Language (SHACL)\footnote{\url{https://www.w3.org/ns/shacl}}. By presenting the examples alongside the editor, users can easily inspect, run, or adapt the existing queries to their own information needs, or simply formulate completely new queries using the existing ones as source of inspiration. %Based on the methodology in \cite{gigascience-paper}, Listing \ref{lst1} shows an example of how the pairs of text and SPARQL query are serialized in Turtle and accessible via the SPARQL endpoint where the query is executable. 
Figure \ref{fig1} depicts the editor webpage. %The first question on the right side is selected, and its corresponding SPARQL query is rendered. 
By clicking on the ``Browse examples'' button (bottom-right), users can inspect and search the full list of examples. 

\subsection{A context aware autocomplete}\label{sec2_2}

Our SPARQL editor provides a precise autocomplete based on the Vocabulary of Interlinked Datasets (VoID)\footnote{\url{https://www.w3.org/TR/void}} descriptions of any RDF dataset. To generate VoID descriptions, SPARQL endpoint maintainers can rely on the VoID generator tool \cite{gigascience-paper} that automatically produces VoID metadata that include statistics of the given dataset. The maintainers can make these metadata accessible through their respective SPARQL endpoints. The real-world applicability of this tool was demonstrated by running it in very large RDF datasets such as the UniProt SPARQL endpoint\footnote{\url{https://sparql.uniprot.org}}, which contains more than 210 billion triples. 
 
Our editor retrieves the VoID metadata, if available, by querying directly the endpoint itself. The main information of interest processed by the editor from the VoID descriptions is represented by the classes instantiated in the endpoint and properties that relate them. %The VoID description is directly extracted from the SPARQL endpoint that is used by our extended \textit{Yasgui} environment, thus accurately reflecting the actual data. 
This ensures that only properties that are stated and applicable according to the schema and the known type of a variable are suggested in the autocomplete drop-down menu (see Figure \ref{fig1}). If no VoID metadata are found, the editor will try to retrieve all classes and properties.
%For further details on the VoID description generator used, see the corresponding GitHub repository \cite{}. 
Finally, our autocomplete approach is lighter, faster, and complementary when compared to the state-of-the-art approaches such as the Qlever autocomplete. In addition, it works within SPARQL 1.1 federated \texttt{SERVICE} calls.

% As an additional convenient feature, the RDF prefixes defined as metadata with the SHACL property \texttt{sh:prefixes} %such as in the description of the SPARQL query examples
% are used by the editor to declare prefixes automatically when writing or editing queries. For example, by typing \texttt{rdfs:}, the query editor will add \texttt{PREFIX rdfs: <http://www.w3.org/2000/01/rdf-schema\#>} into the query header. Users can also add all prefixes related to a given SPARQL endpoint by clicking on the ``Add common prefixes" button.

% \vspace{-0.2cm}
\subsection{Building and visualizing a data-aware schema} 
% \vspace{-0.2cm}

By also considering the VoID descriptions, %see Subsection \ref{sec2_2} for more information,
our query editor is able to build and visualize a simplified data schema that reflects exactly the RDF properties and classes used and accessible through a given SPARQL endpoint. %This helps to visualize which properties and classes are stated in an endpoint. As a result, our SPARQL editor mitigates the issue of RDF datasets that lack of a more informative data schema. %In practice, RDF datasets may or may not explicitly implement all properties and classes described in the Web Ontology Language (OWL) ontologies they utilize. %This is also because ontologies operate under the open-world assumption. 
Therefore, a data schema based on the actual contents of the endpoint, that is, a \textit{data-aware schema}, is useful to aid users in writing SPARQL queries. To display the data-aware schema of a given endpoint, users can click on the ``Classes overview'' button in our graphical query editor interface (see Figure \ref{fig1}, top orange buttons).

\section{Demonstration}
The proposed SPARQL query editor is publicly available and deployed at \url{https://sib-swiss.github.io/sparql-editor}. Users can choose either a predefined endpoint shown in the SPARQL endpoint field at the top, or input their own endpoint of interest in this field. Note that the new features described in Section \ref{sec2} will only be available if the corresponding metadata are accessible through the assigned SPARQL endpoint. At the time of writing, 8 out of 10 public endpoints listed in our official query editor release included all the metadata needed for our tool to be fully operational. This also demonstrates that our approach is triple store agnostic, and readily deployable over diverse RDF data sources. Finally, non-savvy SPARQL users can use our ExpasyGPT tool at \url{https://www.expasy.org/chat} that generates SPARQL queries from plain-English text and run these queries with our SPARQL editor.

\section{Limitations}

\begin{itemize}

\item \textbf{Editor features limitations:} YASGUI relies on CodeMirror 5\footnote{\url{https://codemirror.net/5/}} which is outdated and does not have all the latest features of CodeMirror 6. We investigated the later one, but we realised that upgrading it would require to rewrite several core functionalities, such as autocomplete. Alternatively, we also consider replacing the editor with a VisualStudio based editor and a language server protocol (LSP) for SPARQL, for example, the one developed by the Qlever team\footnote{\url{https://github.com/IoannisNezis/Qlue-ls}}.

\item \textbf{Property autocomplete:} For property autocomplete to fully work, users need to explicitly define the subject type. This could be improved by reusing methods from the efficient and effective SPARQL autocompletion paper\cite{bast2022efficient}.

\item \textbf{Target users:} This tool is intended for users that knows the SPARQL query language. Nevertheless, writing a query often requires exploratory steps to understand the data schema, that is how the data are structured. Our editor aims to reduce this friction and assist users in dealing with complex or unfamiliar endpoints.

% One of the limiting factors in the adoption of SPARQL is the lack of helpful tools to write queries.
% One of the barriers to SPARQL adoption is the lack of supportive tools.

\item \textbf{Visual query builder:} Supporting less tech-savvy users through visual query building is currently out of scope; we recognize its value and believe our work on exposing and standardizing endpoint metadata could serve as a foundation for such tools. We welcome future contributions or collaborations that aim to expand the tool in this direction.

\end{itemize}

% A factor limiting SPARQL usage is the lack of helpful tool to write SPARQL queries. Writing a SPARQL query usually requires many exploratory queries to understand the structure of the data, an editor with relevant autocomplete would help a lot.

% Our current focus is on providing an advanced SPARQL editor tailored to users who are already familiar with query writing, offering popular features such as linting, error highlighting, and intelligent autocompletion based on endpoint metadata. While supporting less tech-savvy users through visual query building is currently out of scope, we recognize its value and believe our work on exposing and standardizing endpoint metadata could serve as a foundation for such tools. We welcome future contributions or collaborations aimed at expanding the tool’s accessibility in that direction.

\section{Conclusion}
We have presented an easy-to-deploy, open source query editor over any SPARQL endpoint, which includes several novel features powered by lightweight metadata. %The editor was mainly implemented as an extension of the \textit{Yasgui} tool. 
The proposed features can greatly help SPARQL-savvy users to write queries over any SPARQL endpoint of interest. We also successfully applied our query editor to several public SPARQL endpoints that range from small to very large real-word RDF datasets. As a future work, we aim at improving the autocomplete feature to suggest next triple patterns by also considering predefined functions and rules. Currently, the proposed features will not work properly, if the needed metadata are not provided. To mitigate the later issue, we will work on an approach that probes endpoints and RDF data dumps, if available. Finally, we welcome contributions from the community to develop this further in our GitHub repository. 

\subsubsection{Acknowledgements} This work was funded by the Swiss National Science Foundation (SNSF) under the CHIST-ERA-22-ORD/SNSF TRIPLE project, grant: 20CH21\_217482; the SNSF MetaboLinkAI project, grant: 10.002.786; and the swissuniversities CHORD in Open Science I, grant: “Swiss DBGI-KM”.

%
% ---- Bibliography ----
%
% BibTeX users should specify bibliography style 'splncs04'.
% References will then be sorted and formatted in the correct style.
%
\bibliographystyle{splncs04}
\bibliography{refs}

\end{document}